**Title**: Identifying melancholic depression biomarker using whole-brain functional connectivity.

**Authors:**
Naho Ichikawa, Giuseppe Lisi, Noriaki Yahata, Go Okada, Masahiro Takamura, Makiko Yamada, Tetsuya Suhara, Ryu-ichiro Hashimoto, Takashi Yamada, Yujiro Yoshihara, Hidehiko Takahashi, Kiyoto Kasai, Nobumasa Kato, Shigeto Yamawaki, Mitsuo Kawato, Jun Morimoto, Yasumasa Okamoto.

[Abstract]

**Importance**

Melancholia, with biological homogeneity, has been a key to understand critical features of major depressive disorder (MDD). Previous biomarker studies have focused on severe treatment-resistant depression with recurrent episodes. However, for effective treatment selection, it is also very important to focus on majority of depression like melancholia, from general clinic outpatients at early stages of depression.

**Objective**

By focusing on melancholic features with biological homogeneity, this study aimed to identify a small number of critical functional connections (FCs) that were specific only to the melancholic type of MDD.

**Design, Setting, and Participants**

On the resting-state fMRI data, classifiers were developed to differentiate MDD patients from healthy controls (HCs). A completely independent validation cohort was prepared. We examined correlations between the biomarker score, which is the value to predict the liability of melancholic MDD (i.e., weighted linear sum) and depression symptoms (i.e., the Beck Depression Inventory scores). In addition, this biomarker was applied to the data of pharmacological treatments and of other mental disorders.

**Results**

The classification accuracy was improved from 50 % (93 MDD and 93 HCs) to 70% (66 melancholic MDD and 66 HCs), when we specifically focused on the melancholic MDD with moderate or severer level of depressive symptoms. It showed 65% accuracy for the independent validation cohort. The biomarker score distribution showed improvements with escitalopram treatments, and also showed significant correlations with depression symptom scores (BDI, $r = .59$). This classifier was specific to melancholic MDD, and it did not generalize in other mental disorders including autism spectrum disorder (ASD, 54% accuracy) and schizophrenia spectrum disorder (SSD, 45% accuracy). Among the identified 12 FCs from 9,316 FCs between whole brain anatomical node pairs, the left DLPFC / IFG region, which has most commonly been targeted for depression treatments, and its functional connections between Precuneus / PCC, and between right DLPFC / SMA areas had the highest contributions.

**Conclusions and relevance**

Given the heterogeneity of the MDD, focusing on the melancholic features is the key to achieve high classification accuracy. The identified FCs specifically predicted the melancholic MDD and associated with subjective depressive symptoms. These results suggested key FCs of melancholic depression, and open doors to novel treatments targeting these regions in the future.

# Introduction

For a typical clinical diagnosis, major depressive disorder (MDD) has been diagnosed with the categorical criteria of DSM-5. The diagnosis depends on how patients feel or act in several aspects of experiences in daily life, for observable or noticeable symptoms. However, Research Domain Criteria (RDoC) framework assumes that mental disorders can be addressed as disorders of brain circuits (Insel, Cuthbert, Garvey et al., 2010).

In order to identify robust functional characteristics of depressive brains, the resting state functional magnetic resonance imaging (rsfMRI) has been attracted attention as a measure of baseline neural activity with benefit of data-driven analysis based on availability of large datasets, (Cao et al., 2014). According to a recent review, rsfMRI studies in MDD has been using seed-based analysis or/and independent component analysis (Mulders et al., 2015). In addition, a multivariate pattern analysis has shown that the majority of the most discriminating features between MDD and healthy controls were intra- or inter-connectivity of the default mode network (DMN; Zeng, et al., 2012). For these data-driven analyses for depression biomarkers above, though those studies showed comparatively high sensitivity (75-92%) and specificity (75-88%), those accuracies were not for an independent validation cohort. When affected by idiosyncratic characteristics of a particular sample (e.g., local site, MRI scanner type, and recruit process, etc.), regression models will tend to cause overfitting. Some cautions and recommendations were provided to be careful about inflated predictions of psychiatric neuroimaging (Whelan and Garavan, 2014). The other study focused on the generalization to a diverse sample, which included 360 depression and 360 healthy controls in total, reported rather lower number of classification accuracies around 45 – 56 % (i.e., chance level was 50 %) in the large heterogeneous sample, and significantly higher accuracy than chance level in a subgroup with higher depression severity, 61 – 62 % for an independent validation (Sundermann et al., 2016). In order to overcome difficulties of generalization, Drysdale et al. (2016) applied the combined algorithm of hierarchical clustering and SVM to the training dataset which includes only treatment-resistant MDD patients (TRD; with a history of failure to respond to at least two anti-depressant medication trials). Because their hierarchical clustering based only on the TRD subjects, it might be more useful as a TRD biomarker. For more general clinical assessment of depression, it would be important to develop a clinic-based biomarker for MDD including those who are at early stages of depression.

In the fields of psychiatry, melancholia has been known as a subtype determined by strong biological factors (Parker et al., 1996). Melancholia has been characterized with an increased risk of MDD in co-twins, which was greater in monozygotic than dizygotic pairs; the symptoms of higher rates of comorbidity with anxiety disorder; overall severity; larger number of episodes; and lower levels of neuroticism (Kendler,1997; Sun, 2012). It has also been reported that the melancholia is more responsive to pharmacological treatment compared to non-melancholic MDD (Parker et al., 1996). As a genetic biomarker of depression, the CONVERGE consortium has successfully identified two loci by focusing on a severe subtype, melancholia feature with recurrent MDD episodes (CONVERGE consortium, 2015). As a main subtype with considerable amount of previous studies, focusing on the melancholia is one of the traditional approaches to understand depression. Hyett et al. (2015) observed reduced effective connectivity during rsfMRI involving in attention and interoception in melancholia, compared to non-melancholic depressive and healthy control groups, though with comparatively small sample size per group.

Recently, Yahata et al. (2016) proposed a machine-learning based algorithm that demonstrated a high degree of generalization in classification performance for multiple independent validation cohorts of autism spectrum disorder (ASD). Their sparse algorithm selected only 16 out of all 9,730 FCs that distinguished individuals with ASD from typically-developed individuals with accuracy of 85%. The reliability of the classifier was generalized in an independent validation cohort of ASD from overseas (75% accuracy). They also found that the ASD classifier was specific to the ASD.

In this study, we applied this algorithm to form a classifier for the MDD patients from general clinics with no treatment or at beginning of treatment. The purpose of this study is: (1) to examine if we could have higher classification accuracy by limiting the training dataset from all MDD outpatients to the subtype of melancholia with moderate or severe depressive symptoms; (2) to test if we could generalize the results of (1) to independent validation cohorts; (3) to examine the association between the selected FCs and severity of depressive symptoms, by using the BDI-II scores (Beck Depression Inventory; Beck et al., 1996) and by comparing the classification results between pre- and post- escitalopram treatment (for 6 weeks) datasets; and (4) to clarify the specificity of depression classifier by applying to the independent datasets of other mental disorders, including ASD and schizophrenia spectrum disorder (SSD).

## Methods
### Participant Selection for the training dataset
All the patients were recruited at the Hiroshima University Hospital and local clinics (in Hiroshima, Japan) and screened using the M.I.N.I. (Sheehan et al. 1998; Otsubo et al. 2005) for a MDD diagnosis with the DSM-IV criteria. Exclusion criteria included current or past manic episodes; psychotic episodes; alcohol dependence or/and abuse; substance dependence or/and abuse; and antisocial personality disorder. Patients had initial MRI scan before or after starting medication within 0-2 weeks. Healthy participants were recruited from local community, interviewed with the M.I.N.I., and none showed any history of psychiatric disorders. The study protocol in this study was approved by the Ethics Committee of Hiroshima University. Prior to the administration of any experimental procedure, written informed consents were obtained from all participants. For the training dataset of the all MDD classifier, first, we used all the melancholic and non-melancholic MDD data collected from four different sites to see if it works for the entire heterogeneous depression cohort. Then, for the melancholic MDD classifier, the training dataset was limited to have the subtype of melancholia (based on M.I.N.I.) with moderate depression symptoms for patients, with age- and gender-matched healthy controls, based on the Beck Depression Inventory (Beck et al., 1996; BDI-II score 17 or higher). The numbers of patients and healthy controls were set to be equal, in order to develop a classifier unbiased toward either group (See Table 1). For the scores of the Japanese version of the national adult reading test (JART; Matsuoka et al., 2006), which was used to estimate the intelligence quotient (IQ), there were three missing data in the training datasets (i.e., one melancholic MDD and two HCs in both all MDD and melancholic MDD datasets) and two missing data of treatment-resistant MDD in the test dataset. For BDI scores, there were two missing data of HCs (only in the all MDD dataset) and one missing post-antidepressant treatment data in the training dataset.

### Generalization to an independent external validation cohort
An independent validation cohort was formed at the National Institute of Radiological Sciences in Japan. The participants were evaluated on lifetime history of psychiatric disorders based on M.I.N.I. The MDD patients were free of comorbid psychiatric disorders, and all healthy controls were free of any somatic, neurological, or psychiatric disorders and had no history of current or previous drug abuse. All participants provided written informed consent before the study. The protocol was approved by the Radiation Drug Safety Committee and by the institutional review board of the National Institute of Radiological Sciences, in accordance with the ethical standards laid down in the 1964 Declaration of Helsinki and its later amendments.

### Association with measures of depression and depressive symptoms
The Beck Depression Inventory (BDI) is one of the most frequently used instruments for measuring depression and depressive symptoms. It was examined if this biomarker would be associated with severity of depressive symptoms in a whole sample including all MDD as a state measure.

### Pharmacological treatment effects
Twenty-eight patients with melancholic depression in the training dataset had an additional MRI scan after 6 weeks of treatment with antidepressant (escitalopram). We applied the classifier to this post-treatment dataset to examine if it would be a state marker of depression severity and would be sensitive to changes associated with the pharmacological treatments.

### Application to non-melancholic MDD, treatment-resistant MDD
In order to make sure if this biomarker would be specific to the characteristics of melancholic MDD, we applied the same classifier to the datasets of non-melancholic and treatment-resistant MDD. Non-melancholic MDDs are from the all MDD dataset with BDI score 17 or higher.

**Application to other mental disorders**
ASD and SSD datasets were adopted from our previous investigation (Yahata et al., 2016). In order to minimize any effect from comorbidity of depression, the ASD dataset was limited to the data with no active antidepressant medication and recorded resting state with eyes-open like all the other datasets.

**Experiment protocol and data acquisition**
The following common instructions and settings were used in all the sites. In the scan room with dimmed lights, participants were asked not to think of anything in particular, not to sleep, and keep looking at a cross mark in the center of the monitor screen. (Details of scan parameters for MRI data acquisition and procedure in each site were shown in Table S3 and Figure S1 in supplementary material.)

**Neuroimaging data preprocessing and interregional correlations**
All the rsfMRI data was preprocessed using the identical procedures described in Yahata et al. (2016). T1-weighted structural image and resting state functional images were preprocessed using SPM8 (Wellcome Trust Centre for Neuroimaging, University College London, UK) on Matlab R2014a (Mathworks inc., USA). The functional images were preprocessed with slice-timing correction and realignment to the mean image. Then, using the normalization parameters obtained through the segmentation of the structural image aligned with the mean functional image, the fMRI data was normalized and resampled in 2 x 2 x 2 mm$^3$ voxels. Finally, the functional images were smoothed with an isotropic 6mm full-width half-maximum Gaussian kernel. After these preprocessing steps, the scrubbing procedure (Power JD et al. 2012) was performed to exclude any volume (i.e., functional image) with excessive head motions, based on the frame-to-frame relative changes in time series data. In order to keep data quality high enough the for subsequent analyses, we only included the data with more than 50% of the volumes survived in the time series. (For a summary of head motion, see Table S2 in supplementary material.)
For each individual, the time course of fMRI data was extracted for each of 137 regions of interests (ROIs), anatomically defined in the Brainvisa Sulci Atlas (BSA; http://brainvisa. Info; Perrot et al., 2011; Riviere et al., 2002) covering the entire cerebral cortex. In the present study, we did not incorporate the cerebellum in the construction of a classifier, because for many participants in site 1, the cerebellum was truncated in their structural and functional images. After applying a band-pass filter (0.008 - 0.1 Hz), the following nine parameters were linearly regressed out: the six head motion parameters from realignment; the temporal fluctuation of the white matter; that of the cerebrospinal fluid; and that of the entire brain. A pair-wise Pearson correlations between 137 ROIs were calculated to obtain a matrix of 9,316 FCs for each participant.

**Classification algorithm for FC selections**
Here, we applied the identical classification algorithm developed in a previous study on ASD (Yahata et al., 2016), which adopts a cascade of L1-regularized sparse canonical correlation analysis (L1-SCCA) and sparse logistic regression (SLR). SLR has the ability to train a logistic regression model, while objectively pruning FCs that are not useful for the purpose of classifying MDD. Before training SLR, L1-SCCA was used to reduce the input dimension to some extent and simultaneously reduce the effects of nuisance variables (NVs) that may cause catastrophic over-fitting. In this study, site, sex, and age were included in the NVs. The method uses a sequential process of nested-feature selection and leave-one-out cross validation (LOOCV) in order to avoid information leakage and over-optimistic results (Whelan et al. 2014). At the end of LOOCV, the output of the logistic regression classifier was used to compute the classification accuracy, and the associated weighted linear sum (WLS) was used to compute the correlation analysis analysis with the score of depression symptoms (BDI-II). The detailed description of the algorithm is found in the methods section of the ASD paper (Yahata et al., 2016). The original classification code developed for the ASD paper is also available for access (please contact the server administrator of ATR Brain Information Communication Research Laboratory: asd-classifier@atr.jp).

# Results
**Classification performance and generalization**
For the evaluation of classification accuracy, we used the results of LOOCV for the training dataset (See Table 2).

The all MDD sample included 93 patients (both melancholic and non-melancholic; BDI score >= 11) with 93 healthy controls (BDI <= 10). The classifier of the all MDD was composed of 22 FCs as relevant predictors, and the classification accuracy was 51 % (sensitivity 53 %, specificity 48 %, and AUC 0.52). The classification accuracy was just around the chance level (50 %), when all heterogeneous MDD samples were included. Then, we limited samples to focus only on the subtype of melancholia with moderate or higher depression symptoms (BDI >=17). The second training dataset included 66 melancholic MDD patients and 66 healthy controls (BDI<=10). The melancholic MDD classifier selected 12 FCs, and its LOOCV classification accuracy was 70 % (sensitivity 64 %, specificity 77 %, and AUC 0.77; see Figure S3 for more detailed results by site). After running the permutation test (1,000 repetitions) to make sure that the classification accuracy is significant (p<.05; for the permutation results, see Figure S2 in supplement), we checked the performance of generalization in the test datasets as independent validation cohorts. The classification accuracy of Test1 dataset was 61% including all MDDs, but when the patient group was limited only to have the melancholic MDD, the accuracy level was improved to 65% (sensitivity 64 %, specificity 65 %, and AUC 0.62). The results of permutation test on the test data was also significant (p<.05).

**Identified 12 FCs in the melancholic MDD classifier**
All the identified twelve FCs for the melancholic MDD classifier were sorted by their absolute weight, and shown in Figure 1A. Figure 1B shows the absolute value of weight of each FC, as a contribution level to the WLS score of the classifier (for more details, see Table S1 and Figure S4 in supplementary material). The weights of the top two FCs were remarkably high compared to the rest of the FCs. As shown in Figure 1C, FC#1 was the functional connection between
the Left DLPFC (BA46) and the Precuneus / dorsal PCC, whereas the FC#2 was the connection between the left IFG (opecular, BA44) and right DLPFC / supplementary motor area (SMA). Those top two connections had an overlapped ROI regions in left DLPFC, which is often the therapeutic target of transcranial magnetic stimulation (TMS; Fox et al., 2012 for review) for treating depression.

**Change with the escitalopram treatment**                                                                                          A part of
the melancholic MDD in the training dataset went through the program of escitalopram treatment for 6 weeks, and twenty-eight melancholic MDD patients completed post-treatment scans. In addition to the severity of depression symptoms, it was examined if the WLS scores of melancholic MDD classifier would be changed by the escitalopram treatment. Figure 2A shows that the distribution of WLS scores were significantly moved toward that of healthy controls after the escitalopram treatment.

**Prediction of depressive symptom scores (BDI)**
In order to assess if the WLS score of the melancholic MDD classifier works as a predictor of severity of depressive symptoms, we performed a correlation analysis between the WLS and the measured subjective rating scores of BDI for the dataset including all the participants (93 MDDs and 93 healthy controls) and also in all the MDD patients (93 MDDs only). The results showed that there was a significant correlation when all the participants in the training dataset were included ($r = .655$, permutation test $p<.005$, shown in Figure 2B), and when only all the MDD patients included ($r = .188$, permutation test $p<.05$, shown in Figure 2C).

**Application to the test dataset of non-melancholic MDD and TRD**
    The melancholic MDD biomarker did work well for classification neither on the non-melancholic MDD with the accuracy of 54 % (sensitivity 42 %, specificity 67 %, and AUC 0.65), nor on the treatment-resistant MDD with the accuracy of 47 % (sensitivity 40 %, specificity 54 %, and AUC 0.46) (See Figure 3A). However, the distribution of non-melancholic MDD showed a trend to be shifted toward the opposite direction along the axis of the melancholic MDD classifier ($p = .076$, Benjamini–Hochberg-corrected Kolmogorov–Smirnov test).

**Application to the test dataset of other mental disorders**
    The results showed that the classification accuracy of the ASD data was around the chance level, 54 % (sensitivity 55 %, specificity 50 %, and AUC 0.52), and no difference was observed between the distribution of

ASD patients and healthy controls (p=.74, n.s.) The SSD data also showed low accuracy, 45 % (sensitivity 43 %, specificity 47 %, and AUC 0.43). The distribution of SSD also showed a trend to be shifted toward the opposite direction along the axis of the melancholic MDD classifier (p = .057, Benjamini–Hochberg-corrected Kolmogorov–Smirnov test). These results showed that the biomarker developed in this study was specific to the melancholic MDD (See Figure 3B).

**Discussion**

Although it has been difficult to classify the depressed outpatients who visits general clinic for the first time, this study suggested that the classification performance of depression was improved by focusing on the melancholic subtype with moderate or severer symptoms. In addition, the reliability of melancholic MDD classification was generalized to a completely independent validation cohort from a different site. Furthermore, the melancholic MDD biomarker was associated with the severity of depressive symptoms. The distribution of WLS scores in melancholic MDD patients was shifted toward that of healthy controls after the pharmacological treatment with escitalopram. However, this biomarker was specific to the melancholic MDD, and did not work well on non-melancholic MDD, nor treatment-resistant MDD, nor ASD, nor SSD.

Different from previous biomarkers which included only pharmacological treatment-resistant MDD patients to make a depression classifier (e.g., Drysdale et al., 2016), this study included early stage of MDD outpatients in a training dataset as general majority, which has been more difficult to diagnose accurately. It would be clinically meaningful that the moderately high classification performance was obtained from this kind of target cohort, and the reliability of it was generalized to a completely independent cohort with a similar early stage of melancholic MDD patients in a different site. In some previous biomarker studies in genetics or molecular sciences, it has been known that the classification accuracy was improved by focusing on the melancholia. Recently, a biomarker study which used the metabolomics showed improvement from ~ 72 % accuracy for all MDD to 80 % accuracy when focused on the melancholia (Liu et al., 2016). Previous studies on neuroendocrine system also has shown that the melancholic features were associated with larger effect sizes compared to non-melancholic depression (Stetler et al., 2011). Moreover, neurological studies showed consistent differences on cerebrospinal fluid volume gray matter volume, white matter volume, with more evidence of EEG abnormalities during some reward tasks, between melancholic and non-melancholic patients (Parker et al., 2015). Though there are not so many studies on rsfMRI yet, in melancholia, reduced effective connectivity between attention and interoception networks was observed (Hyett et al., 2015). This study supported the idea that patterns of functional connectivity of melancholic and non-melancholic depression patients are different. This suggests that it is important and critical to extract a biologically homogenous group to create a classifier with high performance accuracy.

For the selected 12 features, the top two FCs showed outstanding magnitude of weights compared to the rest of FCs. The overlapped region between these two FCs was in the left dorsolateral prefrontal cortex (DLPFC), which has been a traditional target region of depression treatment by repetitive transcranial magnetic stimulation (rTMS) etc. The brain region around the left DLPFC (BA46/9) and left inferior frontal gyrus (IFG) has been noted with imbalance of left and right DLPFC in MDDs (observed in fMRI and in EEG as alpha asymmetry), that is associated with the negative emotional processing bias (Grimm et al., 2008). The lower left DLPFC activity in MDD patients has also been observed during a verbal fluency task (Okada et al, 2003, Takamura et al., 2016). This region is considered to form the central executive network (CEN; Seeley et al., 2007) and regarded as controlling the default mode based on causal relationships examined by rTMS (Chen et al., 2013).

The rest of selected FCs included many key brain regions which have been reported in previous depression studies. As numerous studies have confirmed, the brain regions in the default mode network (DMN) have been associated with some abnormality or impairment in depression, and the identified FCs included DMN regions as well. The 12 FCs mainly included the brain regions around cingulate cortex, including anterior cingulate cortex (ACC), posterior cingulate cortex (PCC) / Precuneus, thalamus, caudate, left and right DLPFC, visual cortex, and supplementary motor area. Those brain regions were frequently reported in previous studies, showing some structural and/or functional connectivity abnormality in MDD (e.g., Yin et al., 2016). Based on the facts that the WLS scores showed significant correlation with the depressive symptoms, and also the distribution of WLS scores was shifted toward healthy controls after six weeks of escitalopram treatments, this biomarker may be a state marker, rather than a trait marker.

Classification results of ASD and SSD showed that neither of mental disorders was classified well using the melancholic MDD classifier. Furthermore, the other two studies which used the same sparse classification algorithms for developing the ASD classifier (Yahata et al., 2016) and the SSD classifier (Yoshihara, in submission) reported that their FCs had no overlap with the 12 FCs reported in this study for depression biomarker. These results may mean that the main characteristic features and associated neural functional connectivities are different in each mental disorder. On the specificity of the melancholic MDD biomarker, even in the same category of MDD diagnosis, this biomarker did work well neither on non-melancholic MDD nor treatment-resistant MDD.

As limitation, as long as we are using the existing diagnosis, we can only see the results associated with those existing labels but not with any potentially-important unknown factors. This basic question may be solved by having some new perspectives for analysis or by combining unsupervised learning method with combined cross-disorder data together.

**Conclusions**

Our findings suggest that the classification algorithm which was originally developed for ASD was successfully applied to early stages of melancholic MDD outpatients from clinics, to achieve a relatively high classification accuracy and mild level of generalization to independent validation cohorts. Because of heterogeneity of the MDD, it has been very difficult to achieve high classification accuracy when classifiers were developed on all the patients with MDD diagnosis. These results make us realize how important it is to focus on the melancholic features for clinical diagnosis and for biological homogeneity. In addition, the identified functional connectivities were characteristic to the melancholic MDD, but not to other types of MDD and disorders including ASD and SSD. This specific biomarker showed significant correlation with depressive symptoms, and the predicted depression scores in the patient group were improved toward healthies after pharmacological treatments. These results suggest that this may be a state marker rather than a trait marker. Combined with the fact that melancholia has been particularly responsive to biological treatments (Parker et al., 1996), this can be a biomarker of treatment response. Furthermore, as a unique benefit of using the sparse algorithm, only a small number of important FCs with the highest contribution over the whole brain were identified for the biomarker. These identified FCs for the melancholic MDD biomarker can be target brain regions of focused treatment or intervention in the future.

**ARTICLE INFORMATION**
**Author Affiliations:**
Department of Psychiatry and Neurosciences, Graduate School of Biomedical Sciences, Hiroshima University, Hiroshima, Japan; ATR Brain Information Communication Research Laboratory Group, Kyoto, Japan; Department of Molecular Imaging and Theranostics, National Institute of Radiological Sciences, National Institutes for Quantum and Radiological Science and Technology, Chiba, Japan; Department of Youth Mental Health, Graduate School of Medicine, The University of Tokyo, Tokyo, Japan ; Department of Functional Brain Imaging Research, National Institute of Radiological Sciences, National Institutes for Quantum and Radiological Science and Technology, Chiba, Japan; Medical Institute of Developmental Disabilities Research, Showa University, Tokyo, Japan; Department of Language Sciences, Graduate School of Humanities, Tokyo Metropolitan University, Tokyo, Japan; Research Center for Language, Brain and Genetics, Tokyo Metropolitan University, Tokyo, Japan; Department of Psychiatry, Kyoto University Graduate School of Medicine, Kyoto, Japan;

**Corresponding Author:**
Yasumasa Okamoto, MD, PhD
Department of Psychiatry and Neurosciences, Hiroshima University, 1-2-3 Kasumi, Minami-ku, Hiroshima, 734-8551, Japan (oy@hiroshima-u.ac.jp)

**Author Contributions:**
Dr Okamoto had full access to all of the data in the study and takes responsibility for the integrity of the data and the accuracy of the data analysis. Ichikawa and Lisi contributed equally as first authors.

**Conflict of Interest Disclosures:**
None reported.

**Funding / Support:**
This research is conducted as the "Application of DecNef for development of diagnostic and cure system for mental disorders and construction of clinical application bases" of the Strategic Research Program for Brain Sciences from Japan Agency for Medical Research and development, AMED. N.I, G. O., M. T., S. Y., Y. O are also partially supported by "Integrated research on neuropsychiatric disorders" and the Integrated Research on Depression, Dementia and Development Disorders by AMED, Grant Number 16dm0107093h0001.

Table 1. Demographic information of the participants for the melancholic MDD classifier.

**A** Training dataset

|  | All MDD and all controls (Hiroshima) | | Melancholic MDD and matched controls (Hiroshima) | |
|---|---|---|---|---|
|  | MDD | HC | MDD | HC |
| No. of participants | 93 | 93 | 66 | 66 |
| Sex (Male/Female) | 50 / 43 | 44 / 49 | 40 / 26 | 32 / 34 |
| Age, (Mean, SD) | 43.7 (11.9) | 39.3 (11.9) | 43.6 (12.7) | 43.4 (10.2) |
| Severity of depression (BDI-II) | 29.5 (8.6) | 4.1 (3.1) | 30.7 (8.4) | 3.8 (3.1) |
| IQ (JART) | 109.3 (10.3) | 112.6 (8.3) | 107.8 (10.9) | 111.8 (8.5) |
| Previous depressive episodes (single/recurrent) | 0.64 (0.96) | NA | 0.71 (1.03) | NA |
| Melancholia (%) | 74.2 | NA | 100 | NA |
| Comorbidity (Anxiety %) | 2.2 | NA | 2.2 | NA |
| Antidepressant (%) | 47 | NA | 45 | NA |

**B** Test dataset

|  | Independent Cohort All MDD (Chiba) | | Independent Cohort Melancholic MDD (Chiba) | | Antidepressant therapy Melancholic MDD (Hiroshima) | | Non-melancholic (Hiroshima) | | Treatment-resistant (Hiroshima) | |
|---|---|---|---|---|---|---|---|---|---|---|
|  | MDD | HC | MDD | HC | Pre | Post | MDD | HC | MDD | HC |
| No. of participants | 15 | 47 | 11 | 40 | 28 | 28 | 24 | 24 | 25 | 28 |
| Sex (Male/Female) | 9 / 6 | 41 / 6 | 6 / 5 | 35 / 5 | 19 / 9 | 19 / 9 | 10 / 14 | 11 / 13 | 14 / 11 | 12 / 16 |
| Age, (Mean, SD) | 39.7 (10.3) | 24.4 (5.8) | 38.7 (11.5) | 24.1 (4.7) | 43.7 (14.2) | - | 42.7 (9.8) | 31.4 (10.3) | 44.7 (10.0) | 44.4 (8.6) |
| Severity of depression (BDI-II) | 28.8 (10.2) | 4.6 (4.2) | 27.8 (7.5) | 3.7 (3.3) | 29.9 (7.4) | 17.7 (12.3) | 28.1 (7.6) | 4.7 (3.0) | 27.5 (11.8) | 3.3 (3.1) |
| IQ (JART) | NA | NA | NA | NA | 109.4 (11.8) | - | 112.9 (7.8) | 114.0 (8.2) | 110.2 (9.4) | 115.5 (6.1) |
| Previous depressive episodes (single/recurrent) | NA | NA | NA | NA | 0.9 (1.1) | - | 0.4 (0.7) | NA | 0.35 (0.49) | NA |
| Melancholia (%) | 73.3 | NA | 100 | NA | 100 | - | 0 | NA | 52.0 | NA |
| Comorbidity (Anxiety %) | NA | NA | NA | NA | 3.6 | - | 0 | NA | 7.1 | NA |
| Antidepressant (%) | NA | NA | NA | NA | 80 | - | 50 | NA | - | NA |

Table 2. Classification performance of the all MDD biomarker and the melancholic MDD biomarker.

| Data set | AUC | Accuracy (%) | Sensitivity (%) | Specificity (%) | Mean # of Feat. |
|---|---|---|---|---|---|
| [ Training dataset: Discovery cohort (Hiroshima 4 sites) ] | | | | | |
| All MDD (n=93, BDI: 11-53), HC (n=93, BDI:0-10) | 0.52 | 51 | 53 | 48 | 22 |
| Melancholic MDD (n=66, BDI: 17-53), HC (n=66, BDI:0-10) | 0.77 | 70 | 64 | 77 | 12 |
| [ Test dataset 1: Independent cohort (Chiba) ] | | | | | |
| All MDD (n=15, BDI:16-47), HC (n=47, BDI:0-15) | 0.60 | 61 | 60 | 62 | 12 |
| Melancholic MDD (n=11, BDI:17-40), HC (n=40, BDI:0-10) | 0.62 | 65 | 64 | 65 | 12 |
| [ Test dataset 2 : Other types of MDD (Hiroshima) ] | | | | | |
| Non-Melancholic MDD (n=24, BDI:13-42), HC (n=24, BDI:0-10) | 0.65 | 54 | 42 | 67 | 12 |
| Treatment-resistant MDD (n=25, BDI:7-53), HC (n=28, BDI:0-9) | 0.65 | 54 | 42 | 67 | 12 |

The classifier based on the Malancholic MDD dataset was applied to the following Test dataset 1 and 2.

Figure 1. Identified 12 Functional Connectivities for the Melancholic MDD biomarker.

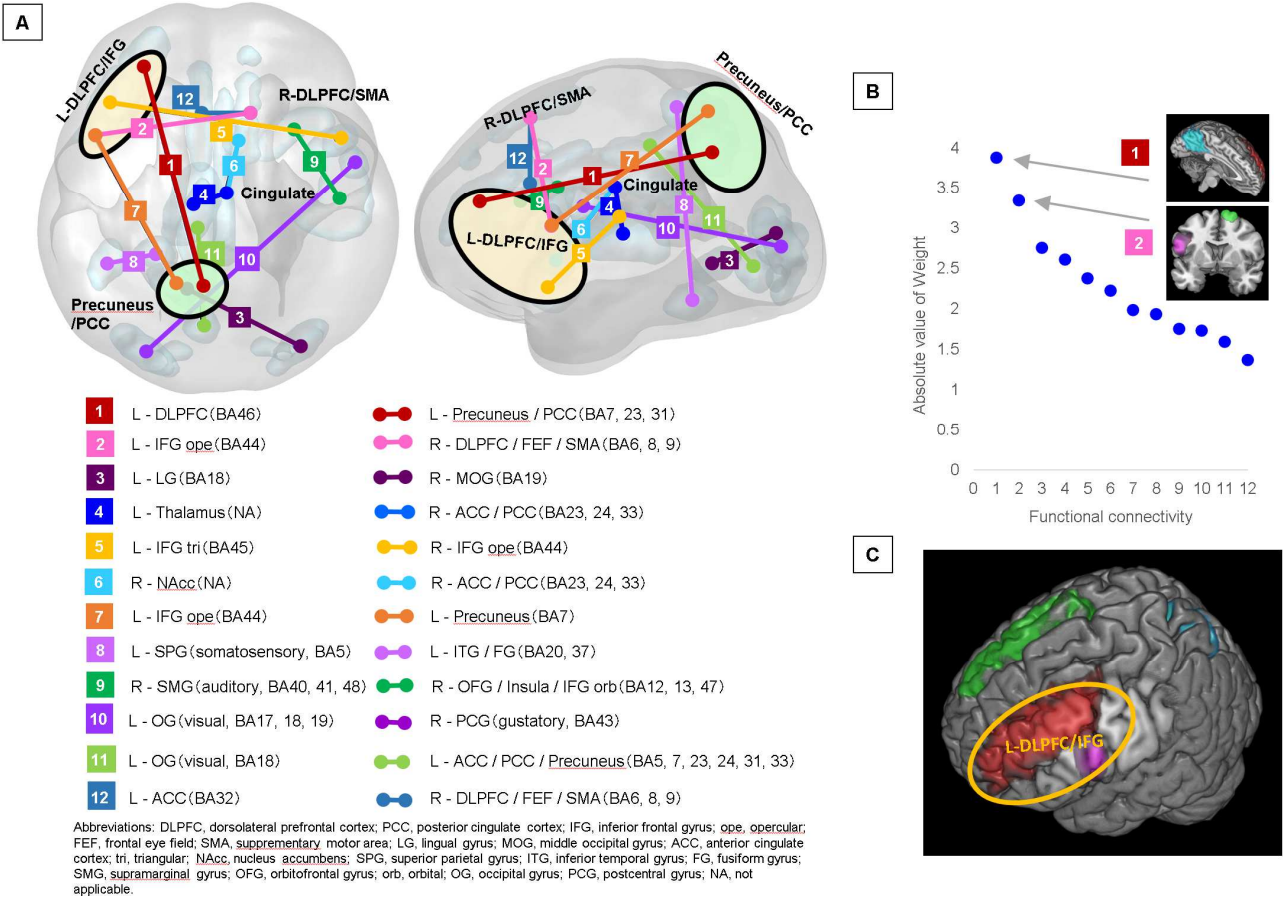

**Figure 2. Associations between the WLS scores of the melancholic MDD classifier and the severity of depression symptoms.**

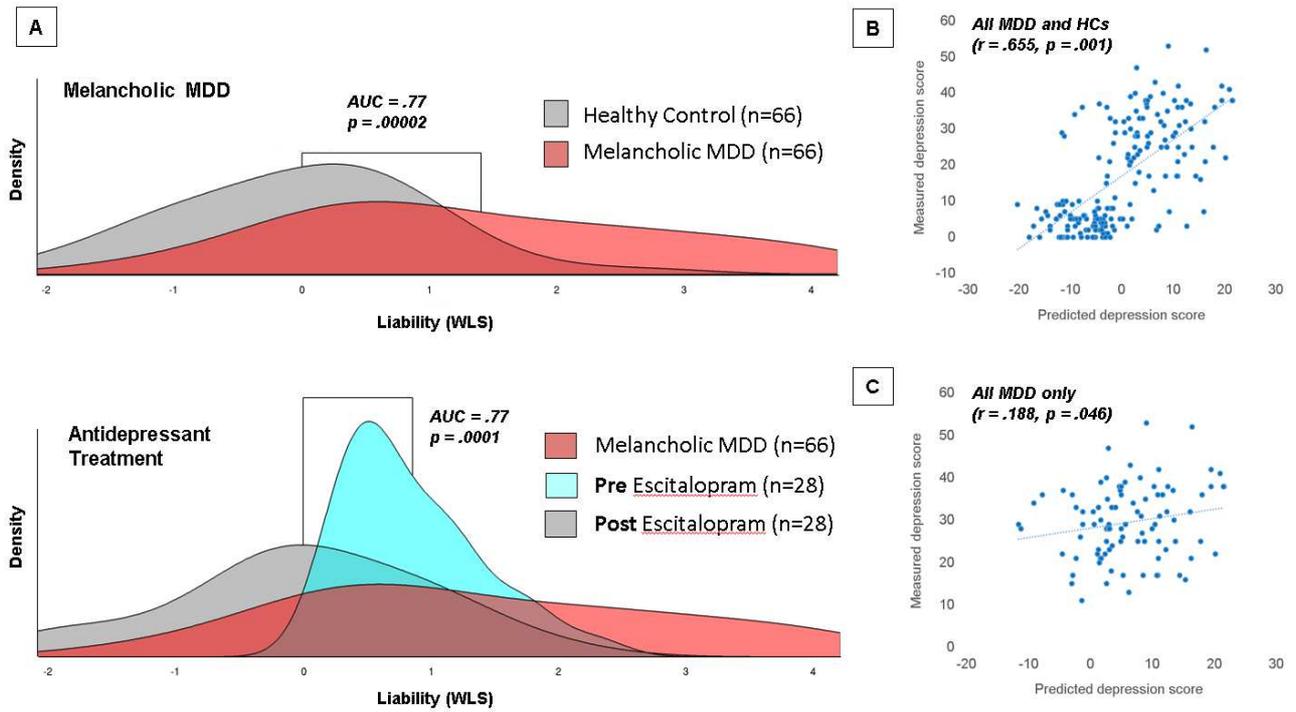

(A) Histograms of the WLS from the training set of the melancholic MDD classifier (upper panel) and the change of the distribution between pre- and post-escitaloplam treatment was significant based on the Benjamini-Hochberg-corrected Kolmogorov-Smirnov test (lower panel). (B) A correlation plot between the WLS and BDI scores in the all MDD training dataset (all MDD: n = 93, HC: n = 93; upper panel), and in only the patient group (all MDD only: n = 93; lower panel).

Figure 3. The specificity of the melancholic MDD classifier to the other affective disorders was examined using the test datasets of MDD subtypes from Non-melancholic MDD and treatment-resistant MDD (TRD) samples (A), and of other disorders from autism spectrum disorder (ASD) and schizophrenia spectrum disorder (SSD) samples (B).

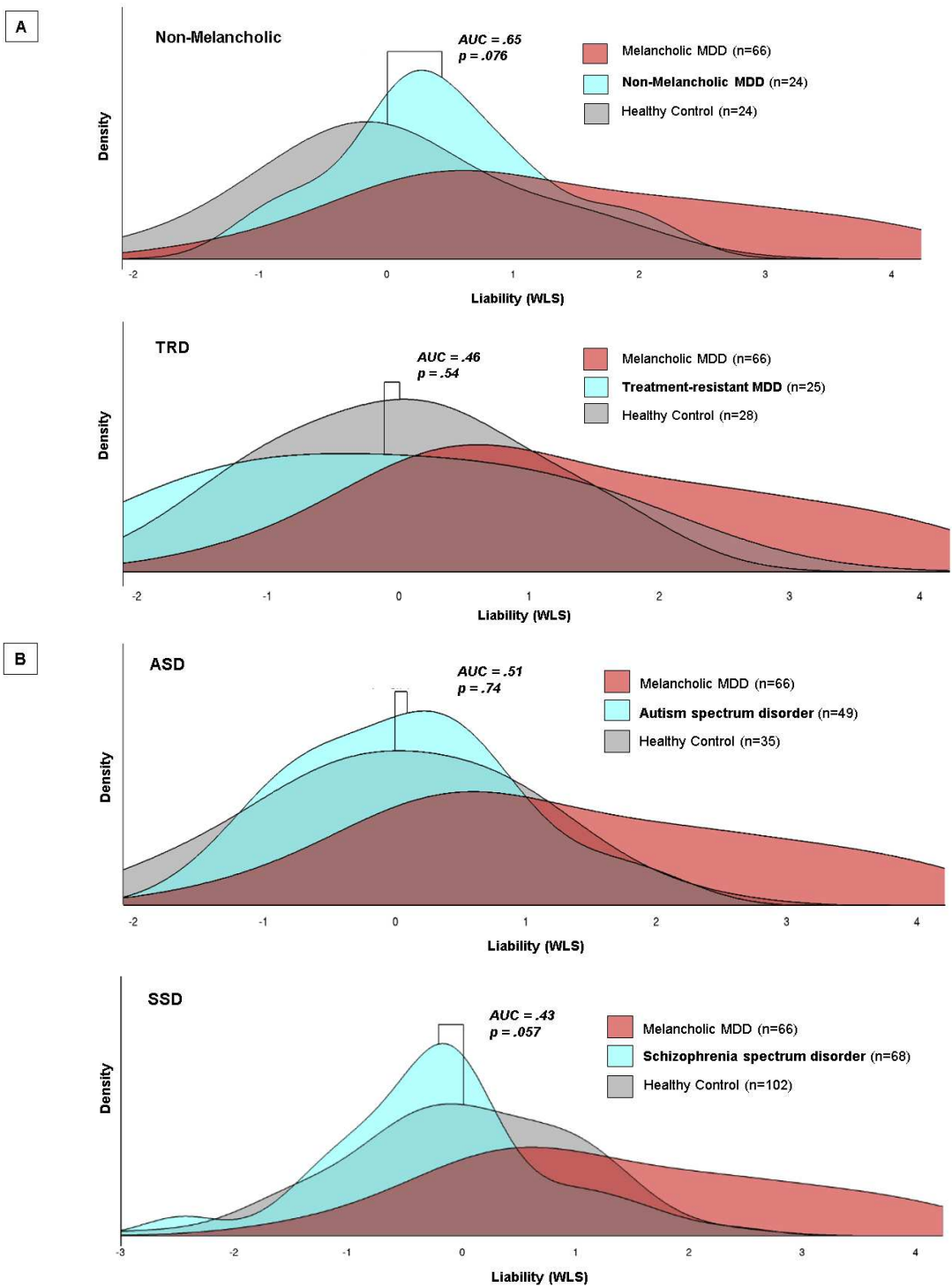

The significance of the Benjamini–Hochberg-corrected Kolmogorov–Smirnov test and AUC values are shown along with the distributions. In this figure, for visualization purposes, the WLS of each data set is standardized to match median and SD of healthy controls across the panels. Note that this WLS standardization is not performed in any quantitative analysis.

Supplementary Material

Table S1. Identified 12 FCs for the the melancholic MDD classifier.

| ID | Name | Lat. | BSA atlas (Sulcus) | BA | Mean FC HC | MDD | Weight |
|---|---|---|---|---|---|---|---|
| 1 | Precuneus / Posterior Cingulate | L | Internal parietal sulcus | 7, 23, 31 | -0.063 | 0.121 | 3.88 |
|  | Middle Frontal Gyrus | L | Intermediate frontal sulcus | 46 |  |  |  |
| 2 | Supplementary Motor (SMA, Pre-SMA), Frontal Eye Fields, Dorsomesial Prefrontal | R | Median frontal sulcus | 6, 8, 9 | 0.175 | -0.017 | -3.34 |
|  | Inferior Frontal Gyrus Opercular | L | Diagonal ramus of the lateral fissure | 44 |  |  |  |
| 3 | Lingual Gyrus | L | Anterior intralingual sulcus | 18 | 0.074 | 0.163 | 2.75 |
|  | Middle Occipital Gyrus | R | Lobe occipital | 19 |  |  |  |
| 4 | Thalamus | L | Thalamus | - | 0.210 | 0.051 | -2.61 |
|  | Posterior Cingulate, Anterior Cingulate | R | Subcallosal sulcus | 23, 24, 33 |  |  |  |
| 5 | Inferior Frontal Gyrus Opecular | R | Inferior precentral sulcus | 44 | 0.408 | 0.288 | -2.38 |
|  | Inferior Frontal Gyrus Triangular | L | Inferior frontal sulcus | 45 |  |  |  |
| 6 | Nucleus Accumbens | R | Accumbens | - | 0.010 | 0.134 | 2.22 |
|  | Posterior Cingulate, Anterior Cingulate | R | Subcallosal sulcus | 23, 24, 33 |  |  |  |
| 7 | Precuneus | L | Superior parietal sulcus | 7 | -0.155 | 0.014 | 1.99 |
|  | Inferior Frontal Gyrus Opercular | L | Diagonal ramus of the lateral fissure | 44 |  |  |  |
| 8 | Superior Parietal Gyrus (Somatosensory) | L | Superior postcentral sulcus | 5 | 0.076 | -0.022 | -1.93 |
|  | Inferior Temporal Gyrus, Fusiform Gyrus | L | Median occipito-temporal lateral sulcus | 20, 37 |  |  |  |
| 9 | Rolandic opercular, Supramarginal Gyrus (Auditory) | R | Posterior lateral fissure | 40, 41, 48 | 0.066 | 0.168 | 1.75 |
|  | Orbitofrontal Gyrus, Insular, Inferior Frontal Gyrus Orbital | R | Anterior lateral fissure | 12, 13, 47 |  |  |  |
| 10 | Postcentral Gyrus (Subcentral) | R | Central sylvian sulcus | 43 | -0.137 | -0.004 | 1.73 |
|  | Occipital Lobe (Visual) | L | Lobe occipital | 17, 18, 19 |  |  |  |
| 11 | Occipital Lobe (Visual Association) | L | Posterior intra-lingual sulcus | 18 | -0.144 | -0.052 | 1.59 |
|  | Anterior Cingulate, Posterior Cingulate, Precuneus (Somatosensory Association) | L | Calloso-marginal posterior fissure | 5, 7, 23, 24, 31, 33 |  |  |  |
| 12 | Supplementary Motor (SMA, Pre-SMA), Frontal Eye Fields, Dorsolateral Prefrontal | R | Median frontal sulcus | 6, 8, 9 | 0.207 | 0.115 | -1.37 |
|  | Anterior Cingulate | L | Calloso-marginal anterior fissure | 32 |  |  |  |

**Supplementary Table S2 | Head Motion of MDD and HC in the training dataset and test dataset.**

|  |  | Training data (Hiroshima) | | | | Test data (Chiba) | | |
|---|---|---|---|---|---|---|---|---|
|  |  | HC | MDD | p | | HC | MDD | p |
| Translation | x | 0.010 ± 0.007 | 0.009 ± 0.007 | 1.550 | | 0.013 ± 0.009 | 0.009 ± 0.005 | 0.919 |
| (in millimeter) | y | 0.041 ± 0.026 | 0.040 ± 0.029 | 5.286 | | 0.045 ± 0.021 | 0.044 ± 0.025 | 5.814 |
|  | z | 0.035 ± 0.019 | 0.029 ± 0.019 | 0.391 | | 0.028 ± 0.014 | 0.017 ± 0.005 | 0.072 |
| Rotation | x | 0.023 ± 0.011 | 0.021 ± 0.012 | 1.290 | | 0.029 ± 0.014 | 0.018 ± 0.005 | 0.127 |
| (in millimeter) | y | 0.010 ± 0.005 | 0.008 ± 0.005 | 0.509 | | 0.011 ± 0.006 | 0.007 ± 0.002 | 0.267 |
|  | z | 0.009 ± 0.005 | 0.007 ± 0.003 | 0.401 | | 0.012 ± 0.010 | 0.008 ± 0.003 | 1.750 |

\* Head radius is calculated as 50mm.

**Supplementary Table S3 | Scanner information and resting-state fMRI protocols of MDD, HC, autistic spectrum disorder (ASD), and Schizophrenia (SCZ).**

| Parameter | Training data | | | |
|---|---|---|---|---|
| | Site1 | Site2 | Site3 | Site4 |
| Participants (Patients/HC) | 57 / 66 | 8 / 47 | 23 / 29 | 17 / 3 |
| MRI scanner | GE Signa HDxt | GE Signa HDxt | Siemens Magnetom | Siemens Verio |
| Magnetic field (T) | 3.0 | 3.0 | 3.0 | 3.0 |
| Field of view (mm) | 256 | 256 | 192 | 212 |
| Matrix | 64 × 64 | 64 × 64 | 64 × 64 | 64 × 64 |
| Number of slices | 32 | 32 | 38 | 40 |
| Number of volumes | 150 | 150 | 112 | 244 |
| In-plane resolution (mm) | 4.0 × 4.0 | 4.0 × 4.0 | 3.0 × 3.0 | 3.3 × 3.3 |
| Slice thickness (mm) | 4 | 4.0 | 3.0 | 3.2 |
| Slice gap (mm) | 0 | 0 | 0 | 0.8 |
| TR (ms) | 2000 | 2,000 | 2,700 | 2,500 |
| TE (ms) | 27 | 27 | 31 | 30 |
| Total scan time (mm:ss) | 5:00 | 5:00 | 5:03 | 10:10 |
| Flip angle (deg) | 90 | 90 | 90 | 80 |
| Slice acquisition order | Ascending (Interleaved) | Ascending (Interleaved) | Ascending (Interleaved) | Ascending |
| Instructions to participants and other imaging conditions | Please relax. Do not think of anything in particular, do not sleep, but keep looking at the crosshair mark presented. The lights in the scan room were dimmed. | Same as Site 1 | Same as Site 1 | Same as Site 1 |

| Parameter | Test data | ASD | SCZ |
|---|---|---|---|
| | Site5 | Site6 | Site7 |
| Participants (Patients/HC) | 15 / 47 | 49 / 35 | 68 / 102 |
| MRI scanner | Siemens Verio | Philips Achieva / Siemens Magnetom Trio | Siemens Trio / Tim Trio |
| Magnetic field (T) | 3.0 | 3.0 / 3.0 | 3.0 / 3.0 |
| Field of view (mm) | 240 | 224 / 192 | 256 / 212 |
| Matrix | 64 x 64 | 64 x 64 / 64 x 64 | 64 x 48 / 64 x 64 |
| Number of slices | 33 | 45 / 33 | 30 / 40 |
| Number of volumes | 204 | 200 / 150 | 180 / 240 |
| In-plane resolution (mm) | 3.75 x 3.75 | 3.5 x 3.5 / 3.0 x 3.0 | 4.0 x 4.0 / 3.3 x 3.3 |
| Slice thickness (mm) | 3.8 | 3.5 / 3.5 | 4.0 / 3.2 |
| Slice gap (mm) | 0.5 | 0.0 / 0.0 | 0 / 0.8 |
| TR (ms) | 2,000 | 2,500 / 2,000 | 2,000 / 2,500 |
| TE (ms) | 25 | 30 / 30 | 30 / 30 |
| Total scan time (mm:ss) | 6:52 | 8:20 / 5:00 | 6:00 / 10:00 |
| Flip angle (deg) | 90 | 75 / 80 | 90 / 90 |
| Slice acquisition order | Ascending (Interleaved) | Ascending / Ascending (Interleaved) | Ascending / Ascending (Interleaved) |
| Instructions to participants and other imaging conditions | Same as Site 1 | Same as Site 1 | Same as Site 1 |

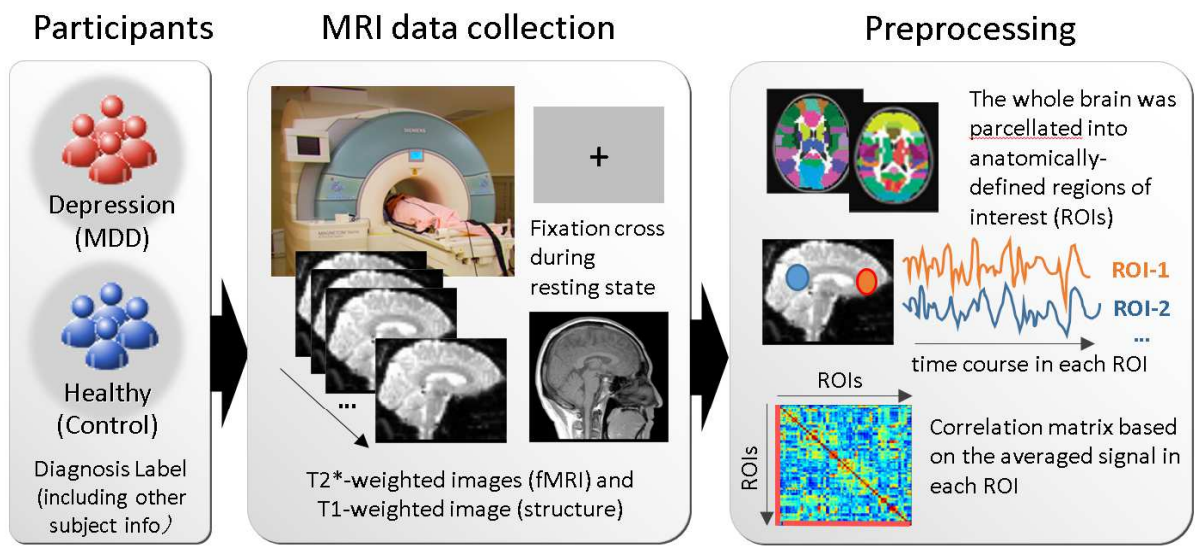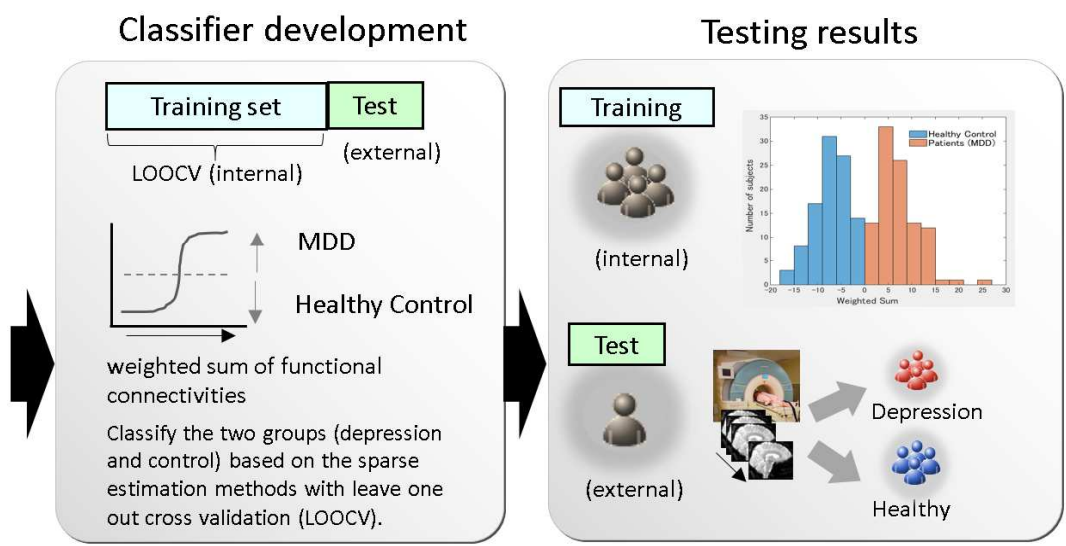

**Supplementary Figure S1. Experimental protocol and biomarker development pipeline**

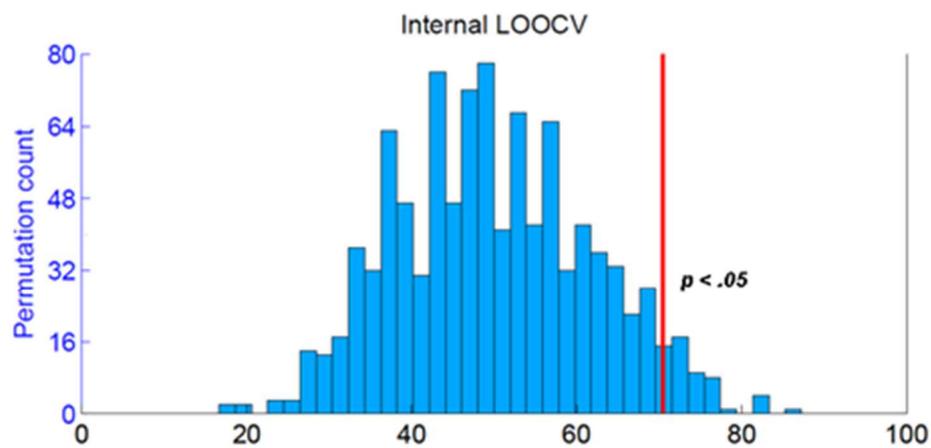
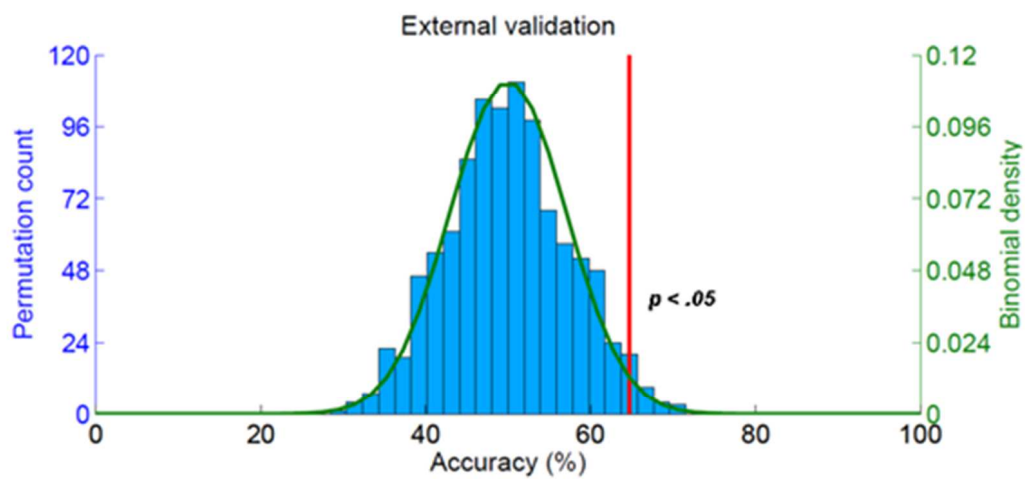

Supplementary Figure S2. Permutation test results.
(a) shows the histogram of the permutation test (1,000 repetitions) for the training dataset LOOCV.
(b) shows the out-of-sample test dataset accuracies, and the binomial distribution is shown as a green curve. The accuracies of the melancholic MDD classifier trained and tested without permutation were shown as red vertical lines. The results of permutation test were significant for LOOCV and for the out-of-sample test (p<.05 for each).

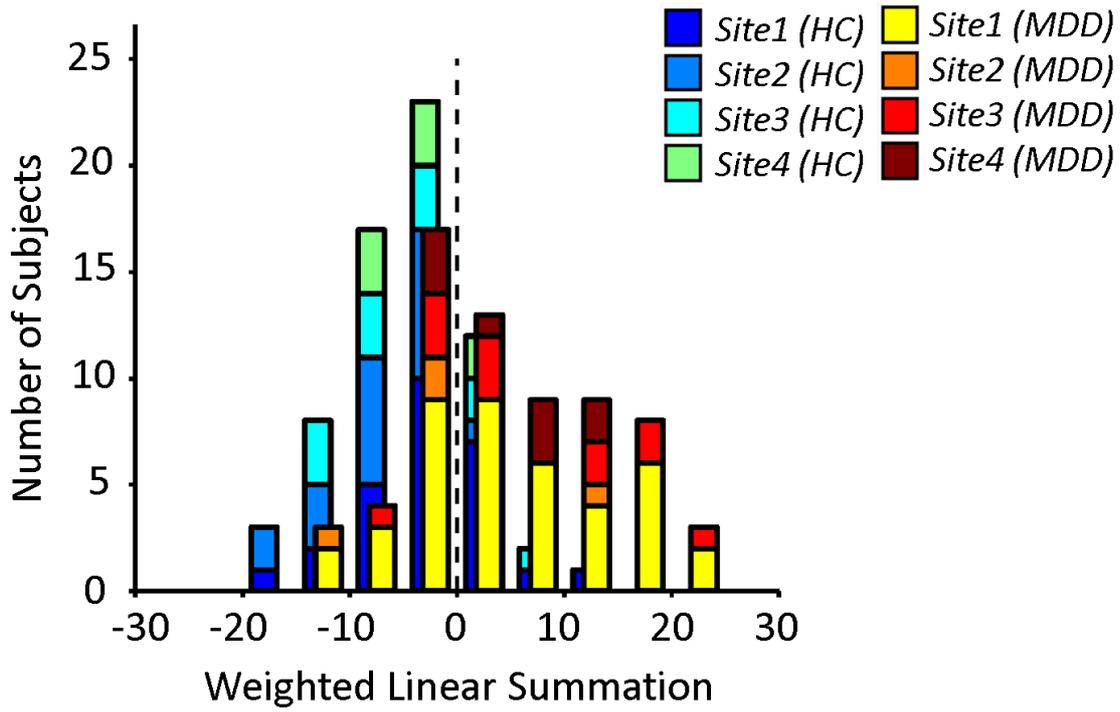

**Supplementary Figure S3.** Weighted linear sum of the training dataset (melancholic MDD: n=66, HC: n=66), colored by site.

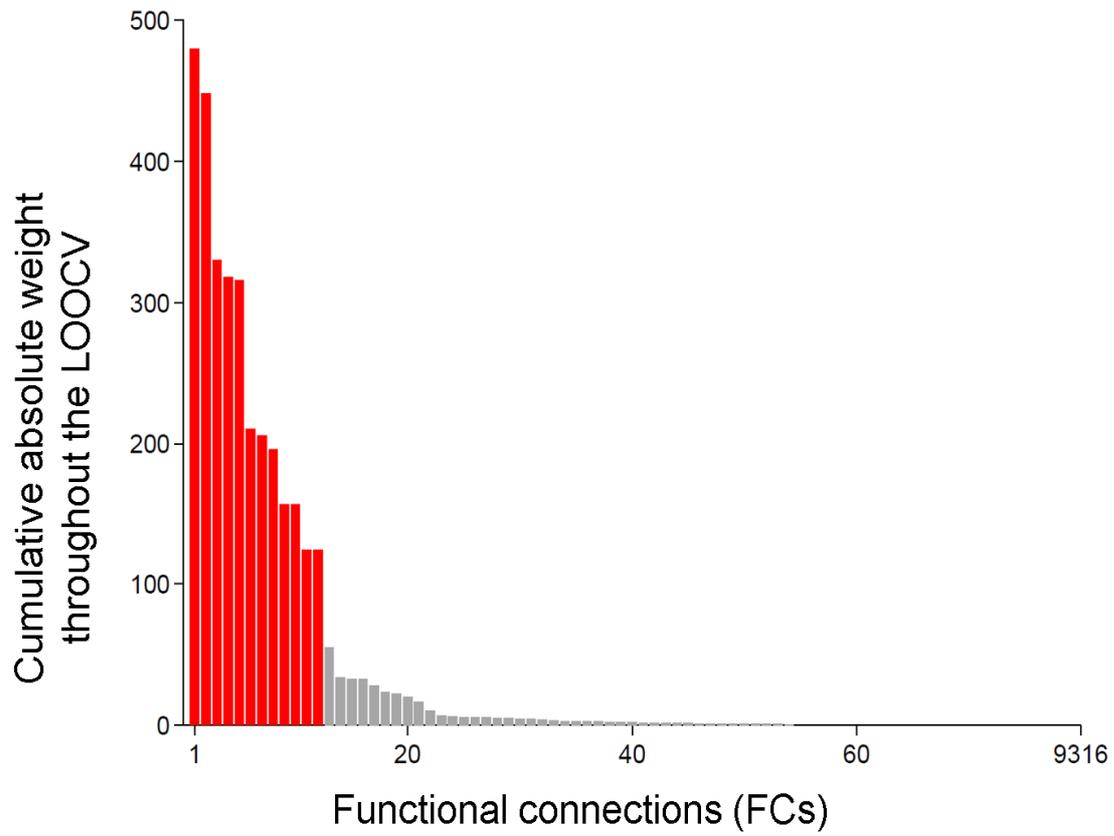

**Supplementary Figure S4. The 12 FCs were identified by the final melancholic MDD - HC classifier.**
From the 54 FCs that were selected at least once throughout the LOOCV. The 12 FCs (in Table 2) are shown in red, and the rest of 42 FCs are in Gray.